# Energy transport faster than light in good conductors and superconductors

Zhong-Yue Wang[1]

*Department of Physics*

*Zhejiang Agriculture & Forestry University*

*No.88 North Ring Road,* Lin'an, *Hangzhou,*

*Zhejiang 311300, China*

**Abstract:** People need a model to study tachyons whose prediction can be tested easily.The dispersion relation $\omega^2 = k^2C^2 - \alpha^2C^2$ of a low-frequency electromagnetic field in good conductors is equivalent to the energy-momentum equation $E^2 = p^2C^2 - m_0^2C^4$ of a tachyon where the proportionality coefficient is $\hbar^2$. An experiment in 1980s to measure the phase velocity $\sqrt{\frac{2\omega}{\mu\sigma}}$ [1] can be regarded as an indirect evidence of the superluminal velocity $V = \frac{1}{\sqrt{\varepsilon\mu}}\left\{\frac{1}{2}\left[\sqrt{1+\frac{\sigma^2}{\varepsilon^2\omega^2}}+1\right]\right\}^{1/2} \approx \frac{1}{\sqrt{\varepsilon\mu}}\left(\frac{\sigma}{2\omega\varepsilon}\right)^{1/2} >> c$ of those photons just equals the rate of energy flow $\frac{\overline{S}}{\overline{w}}$ of the field. Instability of the tachyonic field corresponds to the Joule heat. To detect the speed of energy is difficult and we plan to modulate signals to observe the information velocity (speed of points of non-analyticity) [2].

1. *Corresponding author, E-mail: zhongyuewang(at)yahoo.com.cn*

## I..... Introduction：Rate of energy flow

The velocity $V=\frac{p}{m}$ of a particle is equal to the rate of energy flow $\frac{S}{w}$ [3] in the wave theory

$$V=\frac{p}{m}\xrightarrow{mass-energy equation}\frac{p}{\frac{E}{C^2}}=\frac{C^2}{\frac{E}{p}}\xrightarrow{\frac{S}{g}=C^2\ [4]}\frac{\frac{S}{g}}{\frac{E}{p}}=\frac{\frac{S}{g}}{\frac{w}{g}}=\frac{S}{w} \tag{1}$$

and $\frac{S}{w}>c$ indicates to exceed the light barrier $V>c$ [5].

## II..... Introduction：Low phase velocity

In a dielectric medium, the dispersion relation is $k^2=\omega^2\varepsilon\mu$. As to conductors [6]~[8], the wave vector should be $k+i\alpha$ ( $i=\sqrt{-1}$ ) and

$$(k+i\alpha)^2=\omega^2(\varepsilon+i\frac{\sigma}{\omega})\mu \tag{2}$$

$$k^2-\alpha^2=\omega^2\varepsilon\mu \tag{3}$$

$$k\alpha=\frac{\omega\mu\sigma}{2} \tag{4}$$

Solutions

$$k=\omega\sqrt{\varepsilon\mu}\left\{\frac{1}{2}\left[\sqrt{1+\frac{\sigma^2}{\varepsilon^2\omega^2}}+1\right]\right\}^{\frac{1}{2}} \tag{5}$$

$$\alpha=\omega\sqrt{\varepsilon\mu}\left\{\frac{1}{2}\left[\sqrt{1+\frac{\sigma^2}{\varepsilon^2\omega^2}}-1\right]\right\}^{\frac{1}{2}} \tag{6}$$

Phase velocity $$V_p=\frac{\omega}{k}=\frac{1}{\sqrt{\varepsilon\mu}}\left\{\frac{1}{2}\left[\sqrt{1+\frac{\sigma^2}{\varepsilon^2\omega^2}}+1\right]\right\}^{-\frac{1}{2}} \tag{7}$$

$V_p$ is extremely slow[9]. For commercial power ( $\omega=2\pi\times 60Hz$ ) in copper, $\sigma=5.8\times10^7\Omega^{-1}\cdot m^{-1}$ [9], $\mu\approx\mu_0=4\pi\times10^{-7}\mathrm{H}.m$ [10] (non-ferromagnetic material) and penetration depth $(\frac{\sigma}{\varepsilon\omega} >> 1)$ $\frac{1}{\alpha} \approx \sqrt{\frac{2}{\omega\mu\sigma}} = \sqrt{\frac{2}{2\pi \cdot 60 \times 4\pi \cdot 10^{-7} \times 5.8 \cdot 10^{7}}} = 0.85cm$ [11] (8)

$$k \approx \alpha \qquad \frac{1}{k} \approx \frac{1}{\alpha} = \sqrt{\frac{2}{\omega\mu\sigma}} = 0.85cm \tag{9}$$

$$V_p = \frac{\omega}{k} = 2\pi \cdot 60Hz \times 0.85cm = 3.2m/s \ << c \qquad [12] \tag{10}$$

## III..... de Broglie's theory

In history, de Broglie extended following equations from massless photons to massive matter waves

$$E = mc^2 = \hbar\omega \tag{11}$$

$$p = mV = \hbar k \tag{12}$$

$$V_p = \frac{\omega}{k} = \frac{\hbar\omega}{\hbar k} = \frac{E}{p} = \frac{mc^2}{mV} = \frac{c^2}{V} \tag{13}$$

$$V_p \cdot V = c^2 \tag{14}$$

The phase velocity should be $V_p \geq c$ according to the limit $V \leq c$ of special relativity. Substituting $V_p = 3m/s$ (10) into (14), nevertheless, $V = 10^8 c >> c$ ($c = 3 \times 10^8 m/s$). Two authors inserted metal sheets between the transmitting coil and receiving coil[1]. The phase deviation is consistent with (7).

Actually, $E = mc^2$ and $p = mV$ is valid in both relativity and tachyonics. The difference is that $m = \frac{m_0}{\sqrt{1 - \frac{V^2}{c^2}}}$ for a tardyon ($V < c$) and $m = \frac{m_0}{\sqrt{\frac{V^2}{c^2} - 1}}$ of the tachyon ($V > c$) [13]. Mass $m$ can be eliminated and $V_p \cdot V = c^2$ makes sense in spite of $V_p > c$ or $V_p < c$. A superluminal phase velocity is corresponding to $V < c$. Contrarily, the subluminal $V_p$ implies that $V > c$.

In particular, $V$ will approach infinity for $\omega = 0$ and $V_p = 0$. Unlike ordinary particles, $V$ of a tachyon increases as its energy $E = \frac{m_0 c^2}{\sqrt{\frac{V^2}{c^2} - 1}}$ decreases.

## IV..... Group velocity

The relation given by relativistic mechanics is $V_p \cdot V = c^2$ but people always misunderstand to be

$$V_p \cdot V_g = c^2 \tag{15}$$

which is tenable in circumstances $V = V_g$. That of electromagnetic waves in conductors is a counter-example.

$$\frac{\sigma}{\varepsilon\omega} >> 1 \qquad k = \omega\sqrt{\varepsilon\mu}\left\{\frac{1}{2}\left[\sqrt{1+\frac{\sigma^2}{\varepsilon^2\omega^2}}+1\right]\right\}^{\frac{1}{2}} \approx \sqrt{\frac{\omega\mu\sigma}{2}} \tag{5}$$

$$V_g = \frac{\partial\omega}{\partial k} = 2\sqrt{\frac{2\omega}{\mu\sigma}} = 2V_p\,[14] \xrightarrow{\omega=2\pi\times 60Hz} 6.4m/s << c \tag{16}$$

$$V_p \cdot V_g = 3.2m/s \cdot 6.4m/s << c^2 \tag{17}$$

Hence, group velocities above $c$ are not $V > c$ [15].

## V..... Electrodynamics

Plane waves in a dielectric medium[3],

Energy flow $$\bar{S} = \mathrm{E} \times H = \sqrt{\frac{\varepsilon}{\mu}}\mathrm{E}^2 \tag{18}$$

Energy density $$\bar{w} = \frac{1}{2}(\varepsilon\mathrm{E}^2 + \mu H^2) = \varepsilon\mathrm{E}^2 \tag{19}$$

Speed of energy $$\frac{\bar{S}}{\bar{w}} = \frac{1}{\sqrt{\varepsilon\mu}} \tag{20}$$

In a conductor, there have two answers. One is [16],

Energy flow $$\bar{S} = \frac{k}{2\omega\mu}\left|\mathrm{E}_0\right|^2 e^{-2\alpha x} \tag{21}$$

Energy density $$\overline{w}=\frac{k^2}{2\omega^2\mu}\left|\mathrm{E}_0\right|^2 e^{-2\alpha x} \tag{22}$$

Speed of energy $$\frac{\overline{S}}{\overline{w}}=\frac{\omega}{k}=V_p \xrightarrow{60Hz} 3\,m/s \tag{23}$$

The other is to divide into two kinds [17]

*TM* Energy flow $$\overline{S}=\frac{1}{2\omega\varepsilon}\left|H_0\right|^2 e^{-2\alpha x}\frac{\beta+\dfrac{\sigma}{\omega\varepsilon}\alpha}{1+\dfrac{\sigma^2}{\omega^2\varepsilon^2}}\approx\frac{1}{2\sigma}\alpha\left|H_0\right|^2 e^{-2\alpha x} \tag{24}$$

Energy density $$\overline{w}=\frac{1}{2}\mu\left|H_0\right|^2 e^{-2\alpha x} \tag{25}$$

Speed of energy $$\frac{\overline{S}}{\overline{w}}=\frac{\alpha}{\mu\sigma}=\frac{1}{2}V_p \xrightarrow{60Hz} 6m/s \tag{26}$$

*TE* Energy flow $$\overline{S}=\frac{k}{2\omega\mu}\left|\mathrm{E}_0\right|^2 e^{-2\alpha x} \tag{27}$$

Energy density $$\overline{w}=\frac{\varepsilon}{2}\left|\mathrm{E}_0\right|^2 e^{-2\alpha x} \tag{28}$$

Speed of energy $$\frac{\overline{S}}{\overline{w}}=\frac{k}{\omega\varepsilon\mu} \tag{29}$$

$$V_p\cdot\frac{\overline{S}}{\overline{w}}=\frac{1}{\varepsilon\mu} \qquad \left(V_p=\frac{\omega}{k}\right) \tag{30}$$

The phase velocity $V_p=\dfrac{\omega}{k}$ times speed of energy $\dfrac{\overline{S}}{\overline{w}}$ is $\dfrac{1}{\varepsilon\mu}$. Since $\varepsilon\approx\varepsilon_0$ [9][18][19] and $\mu\approx\mu_0$ [10], it is reasonable to take $\varepsilon_0\mu_0<\varepsilon\mu<10\varepsilon_0\mu_0$ ($c^2=\dfrac{1}{\varepsilon_0\mu_0}$). Owing to $V_p=3m/s$ (10),

$$\frac{\overline{S}}{\overline{w}}=\frac{1/\varepsilon\mu}{V_p}=\frac{0.1c^2\sim c^2}{3m/s}=10^7c\sim10^8c \quad (c=3\times10^8\ m/s) \tag{31}$$

It is of the same order of magnitude of (14).

## VI..... Joule heat

According to Eq.(27),

Change rate
$$\frac{\partial}{\partial x}\bar{S}=-\frac{k\alpha}{\omega\mu}\left|\mathrm{E}_0\right|^2 e^{-2\alpha x} \tag{32}$$

In view of $k\alpha=\frac{\omega\mu\sigma}{2}$ (4),

$$\frac{\partial}{\partial x}\bar{S}=-\frac{1}{2}\sigma\left|\mathrm{E}_0\right|^2 e^{-2\alpha x} \tag{33}$$

On the other hand, the density of Joule's thermal power $I^2R$ $(\frac{I}{\Sigma}=j=\sigma\,\mathrm{E}\,,\ R=\frac{length}{\sigma\Sigma})$ is

$$\sigma\mathrm{EE}=\sigma\frac{\mathrm{Re}(\mathrm{E}^*\,\mathrm{E})}{2}=\frac{1}{2}\sigma\left|\mathrm{E}_0\right|^2 e^{-2\alpha x} \tag{34}$$

Clearly, energy of the electromagnetic field transforms to the Joule heat. $\hbar\omega$ of a single quantum is fixed while strength of the field (density of photons) decreases. In other words, such a tachyonic field is unstable. Entropy(degree of disorder) increases though total energy is conservational.

## VII..... Tachyonics in vacuum and media

The difference of (14) and (30) is related to Abraham-Minkowski controversy[20]~[27]. In the FTL theory [13],

$$E=\frac{m_0c^2}{\sqrt{\frac{V^2}{c^2}-1}} \tag{35}$$

$$p=\frac{m_0V}{\sqrt{\frac{V^2}{c^2}-1}} \tag{36}$$

$$V_p=\frac{\omega}{k}=\frac{\hbar\omega}{\hbar k}=\frac{E}{p}=\frac{mc^2}{mV}=\frac{c^2}{V} \tag{37}$$

$$p^2c^2-m_0^2c^4=E^2 \tag{38}$$

A recent optical experiment shown the constant $c = \frac{1}{\sqrt{\varepsilon_0 \mu_0}}$ is replaced by $C = \frac{1}{\sqrt{\varepsilon \mu}}$ within media[28]. If true, equations of a tachyon should now be

$$E = \frac{m_0 C^2}{\sqrt{\frac{V^2}{C^2} - 1}} \qquad \left( C = \frac{1}{\sqrt{\varepsilon \mu}} \right) \tag{39}$$

$$p = \frac{m_0 V}{\sqrt{\frac{V^2}{C^2} - 1}} \tag{40}$$

$$p^2 C^2 - m_0^2 C^4 = E^2 \tag{41}$$

$$V_p = \frac{\omega}{k} = \frac{\hbar \omega}{\hbar k} = \frac{E}{p} = \frac{mC^2}{mV} = \frac{C^2}{V} \tag{42}$$

$$V_p \cdot V = C^2 = \frac{1}{\varepsilon \mu} \tag{43}$$

They had been utilized to deduce the Cherenkov effect $\cos\theta = \frac{C}{V}$ ( $C = \frac{1}{\sqrt{\varepsilon \mu}} = \frac{c}{n}$ ) of charged particles in media[29] and Eq.(30) is interpreted. Further,

$$k^2 - \alpha^2 = \omega^2 \varepsilon \mu = \frac{\omega^2}{C^2} \tag{3}$$

$$\hbar^2 k^2 - \hbar^2 \alpha^2 = \frac{\hbar^2 \omega^2}{C^2} \tag{44}$$

$$\underbrace{\hbar^2 k^2 C^2}_{p^2 C^2} - \underbrace{\hbar^2 \alpha^2 C^2}_{m_0^2 C^4} = \underbrace{\hbar^2 \omega^2}_{E^2} \tag{45}$$

$$\text{Invariant mass} \quad m_0 = \frac{\hbar \alpha}{C} \tag{46}$$

$$\text{Total energy} \qquad E = \hbar \omega$$

$$\frac{m_0 C^2}{\sqrt{\frac{V^2}{C^2} - 1}} = \hbar \omega \tag{47}$$

$$\frac{\frac{\hbar \alpha}{C} C^2}{\sqrt{\frac{V^2}{C^2} - 1}} = \hbar \omega \tag{48}$$

Velocity of motion $V = C\sqrt{1+\frac{\alpha^2 C^2}{\omega^2}}$ (49)

$$= C\left\{\frac{1}{2}\left[\sqrt{1+\frac{\sigma^2}{\varepsilon^2\omega^2}}+1\right]\right\}^{\frac{1}{2}} \quad (50)$$

In contrast with Eq.(7),

$$V_p.V = C^2 = \frac{1}{\varepsilon\mu} \quad (51)$$

## VIII..... Complex wave vector

Whether the wave vector in de Broglie's wavelength formula might be modulus $\sqrt{k^2+\alpha^2}$ and not the real part $k$? In this case,

$$|\mathrm{k}| = \sqrt{k^2+\alpha^2} = \omega\sqrt{\varepsilon\mu}\sqrt{\frac{1}{2}(\sqrt{1+\frac{\sigma^2}{\varepsilon^2\omega^2}}+1)+\frac{1}{2}(\sqrt{1+\frac{\sigma^2}{\varepsilon^2\omega^2}}-1)} = \omega\sqrt{\varepsilon\mu}\sqrt[4]{1+\frac{\sigma^2}{\varepsilon^2\omega^2}}$$

$$\xrightarrow{\frac{\sigma}{\varepsilon\omega}>>1}\sqrt{\omega\mu\sigma} \quad (52)$$

$$k = \omega\sqrt{\varepsilon\mu}\left\{\frac{1}{2}\left[\sqrt{1+\frac{\sigma^2}{\varepsilon^2\omega^2}}+1\right]\right\}^{1/2} \xrightarrow{\frac{\sigma}{\varepsilon\omega}>>1}\sqrt{\frac{\omega\mu\sigma}{2}} \quad (5)$$

$$|\mathrm{k}| = \sqrt{2}k \quad (53)$$

$$\frac{\omega}{|\mathrm{k}|} = \frac{\sqrt{2}}{2}\frac{\omega}{k} \quad (54)$$

Opposite to that intended, phase velocity is less and $V$ will be bigger. The physical meaning of real part $k$ and imaginary component $\alpha$ is the mechanical momentum $p$ and Compton momentum $m_0C$ respectively. In particle physics, the axial component of an electromagnetic field (simplest gauge field) is zero and photon as a gauge boson is massless ( $m_0 = 0$ ). Such a type called $TEM$ ( $\mathrm{E}_z = 0$, $H_z = 0$ ) waves [30] of electromagnetism occurs in some special cases (free space, infinite dielectric media, coaxial cable,etc). Generally speaking, the longitudinal field is non-zero. For instance, only $TE$ ( $\mathrm{E}_z = 0$ , $H_z \neq 0$ ) or $TM$ ( $\mathrm{E}_z \neq 0$ , $H_z = 0$ ) waves can propagate in a waveguide and $m_0 = \frac{\hbar\omega_c}{c^2} > 0$ [31][5]. The derivation is based on Maxwell's equations having gauge invariance and $m_0$ originates from interactions and boundary conditions. It has nothing to do with Proca's equations to own an inherent non-zero mass even in a free space.

## IX..... Superconductors

Following two views

$\mu \neq 0$ and shielding current (55)

$\mu = 0$ ( $\chi = -1$ ) (56)

are deemed to be equivalent to describe magnetic properties of a superconductor[32][33]. However, the telegrapher's equation

$$\nabla^2 u = \mathcal{L}^* \mathrm{C}^* \frac{\partial^2}{\partial t^2} u \tag{57}$$

is also a wave equation whose phase velocity is $\frac{1}{\sqrt{\mathcal{L}^* \mathrm{C}^*}}$. $\mathcal{L}^*$ and $\mathrm{C}^*$ is inductance and capacitance per unit length. They depend upon permittivity $\varepsilon$ and permeability $\mu$ so $\frac{1}{\sqrt{\mathcal{L}^* \mathrm{C}^*}} \propto \frac{1}{\sqrt{\varepsilon\mu}}$. For example, in a coaxial cable

Inductance
$$\mathcal{L} = \mu \frac{l}{2\pi} \ln \frac{r_1}{r_2} \tag{58}$$

$$\mathcal{L}^* = \frac{\mathcal{L}}{l} = \frac{\mu}{2\pi} \ln \frac{r_1}{r_2} \tag{59}$$

Capacitance
$$\mathrm{C} = \varepsilon \frac{2\pi\, l}{\ln \frac{r_1}{r_2}} \tag{60}$$

$$\mathrm{C}^* = \frac{\mathrm{C}}{l} = \varepsilon \frac{2\pi}{\ln \frac{r_1}{r_2}} \tag{61}$$

Phase velocity
$$\frac{1}{\sqrt{\mathcal{L}^* \mathrm{C}^*}} = \frac{1}{\sqrt{\varepsilon\mu}} \tag{62}$$

Impedance
$$\sqrt{\frac{\mathcal{L}^*}{\mathrm{C}^*}} = \sqrt{\frac{\mu}{\varepsilon}} \frac{\ln \frac{r_1}{r_2}}{2\pi} \tag{63}$$

The magnetic flux $\Phi = \int \mathbf{B}\, d\mathbf{s}$ across a straight bulk superconductor is zero on account of the Meissner effect $\mathbf{B} = 0$ , no matter it is caused by the shielding current or $\mu = 0$ . Consequently, $\mathcal{L} = \frac{\Phi}{I} = 0$ and $\frac{1}{\sqrt{\mathcal{L}^* \mathrm{C}^*}} >> c$. Zero impedance $\sqrt{\frac{\mathcal{L}^*}{\mathrm{C}^*}}$ proportional to $\sqrt{\frac{\mu}{\varepsilon}}$ implies that $\mu = 0$ . Both the phase velocity $\frac{1}{\sqrt{\varepsilon\mu}}$ and rate of energy flow $\frac{1/\varepsilon\mu}{V_p} = \frac{1}{\sqrt{\varepsilon\mu}}$ are infinity. The wavelength should be $\lambda = \frac{V_p}{f} \to \infty$ and phase delay is always zero regardless of the length.

## X..... Mechanism of $\mu = 0$

The Langevin theory can replace phenomenological London equations to explain perfect diamagnetism of superconducting materials.The electric current induced by a magnetic filed $\mathbf{B}$ is

$$I = -e\frac{\omega_L}{2\pi} = -\frac{e^2}{4\pi m}B = -\mu_0\frac{e^2}{4\pi m}H \tag{64}$$

where $\omega_L = \frac{e}{2m}B$ (Larmor frequency). Thus, the magnetic moment is $I\pi r^2$ and magnetisation defined as the moment per unit volume should be

$$M = N \cdot I\pi r^2 = -\frac{\mu_0 Ne^2}{4m}H \cdot r^2 \tag{65}$$

Susceptibility $\chi = \frac{M}{H} = -\frac{\mu_0 Ne^2}{4m}r^2$ is $-1$ and $\mu = (1+\chi)\mu_0 = 0$ of a big atom whose radius is twice the London penetration depth $\sqrt{\frac{m}{\mu_0 Ne^2}}$. Superconductivity should be interdisciphnary of atomic physics and condensed matter physics.

## XI..... Zero-index metamaterials (ZIMs)

Similarly, $\frac{1}{\sqrt{\varepsilon\mu}} \to \infty$ in light of $\varepsilon\mu \approx 0$ [34]~[39]

## Conclusions

Anyway, speed of energy $\frac{\overline{S}}{w}$ of a low-frequency electromagnetic field in conductors equal to the mechanical velocity $V$ of those photons is much faster than $c$ regardless of $k$ or $\sqrt{k^2+\alpha^2}$. It is independent of correctness of the new experiment[28] on account of $V >> c$ is tenable no matter $V_p \cdot V = c^2$ or $V_p \cdot V = C^2$. The Sommerfeld wire[40] is another interesting system to have two types of superluminal electromagnetic fields. Surface wave outside is a little faster than $c$[5] and the speed within the conductor is much greater. By comparison, the dispersion relation $k^2c^2 + \omega_p^2 = \omega^2$ of high-frequency waves is equivalent to relativistic equation $p^2c^2 + m_0^2c^4 = E^2$ ( $E = \frac{m_0c^2}{\sqrt{1-\frac{V^2}{c^2}}} = \hbar\omega$ , $p = \frac{m_0V}{\sqrt{1-\frac{V^2}{c^2}}} = \hbar k$ , $m_0c^2 = \hbar\omega_p$ ). The phase velocity should be $V_p = \frac{\omega}{k} = \frac{c}{\sqrt{1-\frac{\omega_p^2}{\omega^2}}} > c$ and refractive index is $n_p = \frac{c}{V_p} = \sqrt{1-\frac{\omega_p^2}{\omega^2}} < 1$ ( $X$ ray, say). Group velocity $V_g = \frac{\partial\omega}{\partial k} = c\sqrt{1-\frac{\omega_p^2}{\omega^2}} < c$ now comes up to the mechanical velocity $V = c\sqrt{1-\frac{\omega_p^2}{\omega^2}} < c$.

## Nomenclature

$\omega$ = circular frequency

$k$ = wave vector

$\alpha$ = attenuation constant

$\hbar$ = reduced Planck constant ( $\hbar = \frac{h}{2\pi}$ )

$\sigma$ = electric conductivity

$m_0$ = invariant mass of a single particle

$V$ = mechanical velocity of a single particle

$p$ = momentum of a single particle

$E$ = total energy of a single particle

$S$ = energy flow density (Poynting vector)

$w$ = density of energy

$g$ = density of momentum

$\varepsilon$ = permittivity of matter

$\mu$ = permeability of matter

$C$ = a constant dependent on the product of permittivity times permeability

$c$ = light speed in vacuum

$\varepsilon_0$ = permittivity of vacuum

$\mu_0$ = permeability of vacuum

$V_p$ = phase velocity

$V_g$ = group velocity

E = electric intensity

$H$ = magnetic intensity

$B$ = magnetic flux density

$R$ = resistance

$I$ = electric current

$j$ = current density

$\Sigma$ = cross sectional area

$u$ = voltage

$L$ = inductance

C = capacitance

$l$ = length

$r$ = radius

$\chi$ = magnetic susceptibility

$\omega_p$ = plasma frequency

## Postscript: Failure of de Broglie relation in guided optics

| | Dispersion relation | Ratio of $E$ to $p$ deduced from electrodynamics [1] | |
|---|---|---|---|
| Free space | $\frac{\omega}{k}=c$ | $\frac{E}{p}=\frac{w}{g}=c$ | (1) |
| Hollow waveguide | $\frac{\omega}{k}=c$ | $\frac{E}{p}>c$ | (2) |
| Surface wave in vacuum | $\frac{\omega}{k}=c$ | $\frac{E}{p}<c$ | (3) |

Obviously, $E=\hbar\omega$ and $p=\hbar\mathrm{k}$ cannot be simultaneously true otherwise $\frac{E}{p}=\frac{\hbar\omega}{\hbar k}=\frac{\omega}{k}\equiv c$ no matter in free space, waveguides or surface waves. Furthermore, theories and experiments of the Casimir effect [2]~[5] affirm $E\propto\omega=ck$ instead of $E\propto\frac{\omega}{\sqrt{1-\frac{\omega_c^2}{\omega^2}}}$ for photons between two planes which can be regarded as a one-dimensional rectangular waveguide ( $a\to\infty$ or $b\to\infty$ ). Thus, de Broglie's wavelength relation $p=\hbar k$ is not universal. The phase depends on $\frac{pr}{\hbar}$ and not $kr$.

| Evidences of | Travelling waves | Stationary waves |
|---|---|---|
| $p=\frac{h}{\lambda}$ | Compton effect, 1923<br>Davisson-Germer experiment,1927<br>Thomson's experiment,1928, etc. | Bohr-Sommerfeld quantization<br>Density of states |

First two counter-examples to $p=\frac{h}{\lambda}=\hbar k$ in guided optics are now proposed [1]. In practice, the momentum is $p=\hbar\beta$ where the phase constant $\beta$ is just a component of $k$ in the direction of travelling waves ( $\beta<k$ inside a waveguide and $\beta>k$ for surface waves). So-called superluminal

"phase velocity" $\dfrac{\omega}{\beta}=\dfrac{c}{\sqrt{1-\dfrac{\omega_c^2}{\omega^2}}}>c$ and subluminal "group velocity" $\dfrac{\partial\omega}{\partial\beta}=c\sqrt{1-\dfrac{\omega_c^2}{\omega^2}}<c$ in waveguides[6]~[8] is no longer the original definition $V_p=\dfrac{\omega}{k}$ and $V_g=\dfrac{d\omega}{dk}$. The physical meaning of the other component $\sqrt{|k^2-\beta^2|}$ of a photon in guided optics is "Compton momentum" $m_0c$ (the product of invariant mass $m_0$ times light speed $c$) and not "mechanical momentum". As to electromagnetic waves in plasma and conductors, both $E=\hbar\omega$ and $p=\hbar k$ are tenable because the dispersion relation is not limited as $\dfrac{\omega}{k}=c$.